\documentstyle[12pt]{article}

\def\fnote#1#2{\begingroup\def\thefootnote{#1}\footnote{#2}
    \addtocounter{footnote}{-1}\endgroup}

\def\email{makoto@th.phys.titech.ac.jp, JSPS Research Fellow}
\def\sakai{nsakai@th.phys.titech.ac.jp}
\def\sato{sato@th.phys.titech.ac.jp}

\newcommand{\bra}[1]{\mbox{$\langle #1 |$}}
\newcommand{\ket}[1]{\mbox{$| #1 \rangle$}}

\begin{document}

\pagestyle{empty}
\begin{flushright}
	TIT/HEP-323\\
	hep-th/9603173\\
\end{flushright}
\vspace{18pt}

\begin{center}
{\large \bf String scattering off the (2+1)-dimensional} \\ \vspace{4pt}
{\large \bf rotating black hole}

\vspace{16pt}
Makoto Natsuume,\fnote{*}{\email} Norisuke Sakai,\fnote{\dag}{\sakai} and
Masamichi Sato \fnote{\ddag}{\sato}

\vspace{16pt}

{\sl Department of Physics\\
Tokyo Institute of Technology \\
Oh-okayama, Meguro, Tokyo 152, Japan}

\vspace{12pt}
{\bf ABSTRACT}

\end{center}

\begin{minipage}{4.8in}
The $SL(2, R)/Z$ WZW orbifold describes the (2+1)-dimensional black hole which
approaches anti-de~Sitter space asymptotically. We study the $1 \rightarrow 1$
tachyon scattering off the rotating black hole background and calculate the
Hawking temperature using the Bogoliubov transformation.
\end{minipage}

\vfill
\pagebreak

\pagestyle{plain}
\setcounter{page}{1}

\baselineskip=16pt

\section{Introduction}

The Wess-Zumino-Witten (WZW) model is a useful framework to study string theory
in curved spacetime. One simple example is the $SL(2,R)/U(1)$ WZW coset which
is known as Witten Black Hole \cite{witten}. Another simple example is the
$SL(2,R)/Z$ WZW orbifold, which describes a three-dimensional black hole in
asymptotically anti-de~Sitter space \cite{horowitz,kaloper}. The black hole was
originally found as a solution to general relativity by Ba\~{n}ados,
Teitelboim, and Zanelli (BTZ) \cite{btz}, but it was quickly realized that a
slight modification of the solution yields a solution to the bosonic string
theory.

The purpose of this note is to study the $1 \rightarrow 1$ string scattering in
the rotating BTZ black hole background. We solve the tachyon equation in the
linearized approximation and derive the reflection  coefficient for the
scattering. We also derive the Hawking temperature using the Bogoliubov
transformation.

String scatterings in various geometries have been studied in
refs.~\cite{verlinde,sfetsos,ghoroku}. In particular, ref.~\cite{ghoroku}
studies the string scattering in the static ($J=0$) BTZ black hole background.
In the context of the BTZ black hole in general relativity, the scattering of a
massless conformally coupled scalar field has been studied in ref.~\cite{hyun}.

\section{Review}

We will briefly review the BTZ black hole; for more details, see
ref.~\cite{carlip}. The simplest solution of the BTZ black hole is the $(M, J)
= (1, 0)$ case. It is given by
\begin{eqnarray}
ds^{2} & = & 2(k-2) \left\{ -(\hat {r}^2 - 1) dt^{2} + \frac{d\hat {r}^2}{\hat
{r}^2 - 1} + \hat {r}^2 d\hat {\varphi}^2 \right\},
	\nonumber \\
\Phi & = & const.,
	\nonumber \\
H_{\mu \nu \rho} & = & -2(k - 2) \epsilon_{\mu \nu \rho}.
	\label{eq:non-rotating}
\end{eqnarray}
We set $\alpha'=2$. Here, we identify $ \hat {\varphi} \approx \hat {\varphi} +
2 \pi $; this corresponds to a choice of the orbifolding. Starting from a level
$k$ $SL(2, R)$ WZW model, it can be shown that the above solution is an exact
solution to the bosonic string theory. The normalization is obtained as
follows. $k$ appears since we consider the level $k$ WZW model; $-2$ is the
well-known shift of $k$ by the amount of $c_{v}$ \cite{tseytlin}, where $c_{v}$
is the quadratic Casimir of the adjoint representation of the group. The
three-form field strength $H_{\mu \nu \rho}$ is necessary by the Wess-Zumino
term and plays the role of negative cosmological constant. So, the solution
approaches anti-de~Sitter space asymptotically.

Since the central charge of the WZW model is given by $ c = 3k/(k-2) $, $ k =
52/23 $ in order to get $ c=26 $. In the discussion of the BTZ black hole, a
dimensionful parameter $l$ is often used. This parameter can be introduced by
scaling $ \hat {t} $ and $ \hat {r} $. The relation of $l$, the cosmological
constant $\Lambda$, and the level $k$ is given by $ l^2 = - \Lambda^{-1} = 2(k
- 2) $.

An important property of the BTZ black hole is that the general solution is
obtained simply by a different choice of the orbifolding. This property will be
essential in solving the tachyon equation. First, make the transformation
\begin{equation}
\hat {t} = r_{+}t - r_{-}\varphi, \;\;\;
\hat {\varphi} = r_{+}\varphi - r_{-}t, \;\;\;
\hat {r}^2 = \frac{r^{2} - r_{-}^{2}}{r_{+}^{2} - r_{-}^{2}}.
	\label{eq:transf}
\end{equation}
The reason of making this transformation is because we will make the
identification in terms of $ \varphi $ rather than $ \hat {\varphi} $. By the
transformation, the metric (\ref{eq:non-rotating}) becomes
\begin{equation}
ds^{2} = 2(k-2) \left\{ -(r^{2} - M) dt^{2} - J dt d\varphi + r^{2}
d\varphi^{2}
	+ (r^{2} - M + \frac{J^{2}}{4r^{2}})^{-1} dr^{2} \right\}.
	\label{eq:rotating}
\end{equation}
This time we identify $ \varphi \approx \varphi + 2 \pi $ as the orbifolding.
Here, $M = r_{+}^{2} + r_{-}^{2}$; $J = 2r_{+}r_{-}$; and $r_{\pm}$ are the
inner and outer horizons of the black hole.

\section{$1 \rightarrow 1$ Scattering}

Consider the effective action for the tachyon $T$. The spacetime action is
\begin{equation}
S(T) = \int d^{3}X\sqrt{-G}\; e^{-2\Phi} (G^{\mu\nu}\partial_\mu T\partial_\nu
T +
m^2 T^2 + a T^3 +...),
\end{equation}
where $ m^2 = -2$. We expand the tachyon field equation in powers of the
ingoing tachyon. To first order in the tachyon, the field equation is
\begin{equation}
-\frac{1}{e^{-2\Phi}\sqrt{-G}} \partial_{\mu} G^{\mu \nu} e^{-2\Phi} \sqrt{-G}
\partial_{\nu} T + m^2 T = 0.
	\label{eq:tachyon_eq}
\end{equation}
Our task is to solve eq.~(\ref{eq:tachyon_eq}) in the background
(\ref{eq:rotating}). Although $ m^2 = -2 $, we will parametrize $ m^2 = - (4
\lambda^2 + 1)/2(k-2) $. Expressions become simpler in this parametrization.
{}From the actual value of $ m^2 $, $ \lambda^2 = 1/92 $. The parametrization
appears naturally in the representation theory of the $SL(2, R)$ affine
Kac-Moody algebra; the tachyon belongs to the continuous series representation
of the global $SL(2, R)$.

Substituting the metric (\ref{eq:rotating}) into (\ref{eq:tachyon_eq}), we get
\begin{eqnarray}
\lefteqn{(r^{2} - M + \frac{J^{2}}{4r^{2}})\frac{\partial^{2}T}{\partial r^{2}}
+ (3r - \frac{M}{r} - \frac{J^{2}}{4r^{3}})\frac{\partial T}{\partial r}}
	\label{eq:nonzeroJ} \\
& & -(r^{2} - M + \frac{J^{2}}{4r^{2}})^{-1}
\{\frac{\partial^{2}}{\partial t^{2}}
+ \frac{M - r^{2}}{r^{2}}\frac{\partial^{2}}{\partial \varphi^{2}}
+ \frac {J}{r^{2}}\frac{\partial^{2}}{\partial \varphi \partial t}\}T
+ (4\lambda^{2} + 1)T = 0.
	\nonumber
\end{eqnarray}
To solve this equation, expand the field in terms of modes:
\begin{equation}
T(r, t, \varphi) = \sum_{N} \int dE \; T_{E N}(r) e^{-i E t} e^{-iN \varphi}.
	\label{eq:mode}
\end{equation}
Here, $N \in Z$. Eq.~(\ref{eq:nonzeroJ}) is most easily solved by first making
the change of coordinates (\ref{eq:transf}). Then the sum (\ref{eq:mode}) is
rearranged as
\begin{equation}
T(r, \hat {t}, \hat {\varphi})=
\mbox {\large \sf S}_{\hat {N}}\; \mbox {\large \sf S}_{\hat {E}}\; T_{\hat {E}
\hat {N}}(r)e^{-i \hat {E } \hat {t}} e^{-i \hat {N} \hat {\varphi}}.
\end{equation}
Here, {\large \sf S} denotes the summation over the modes $\hat {E}$ and $\hat
{N}$. Also, we set $ E = r_{+} \hat {E} - r_{-} \hat {N} $ and $ N = -r_{-}
\hat {E} + r_{+} \hat {N} $. By the coordinate transformation and the mode
expansion, eq.~(\ref{eq:nonzeroJ}) becomes
\begin{equation}
(\hat {r}^{2} - 1)\frac{d^{2}T_{\hat {E} \hat {N}}}{d \hat {r}^{2}}
+ (3 \hat {r} - \frac{1}{\hat {r}})\frac{dT_{\hat {E} \hat {N}}}{d\hat {r}}
+ ( \frac{\hat {E}^2}{\hat {r}^2-1} - \frac{\hat {N}^2}{\hat {r}^2}
+ 4 \lambda^{2} + 1) T_{\hat {E} \hat {N}} =0.
	\label{eq:zeroJ}
\end{equation}
By changing variables to  $ z = 1 - \hat {r}^2 = \frac{r_{+}^{2} -
r^{2}}{r_{+}^{2} - r_{-}^{2}} $ and
\begin{equation}
T_{\hat {E} \hat {N}}(z) =
z^{i \hat {E}/2} (1 - z)^{i \hat {N}/2} \Psi_{\hat {E} \hat {N}}(z),
\end{equation}
we get the standard hypergeometric differential equation for $\Psi_{\hat {E}
\hat {N}}(z)$:
\begin{equation}
z(1 - z)\frac{d^{2}\Psi_{\hat {E} \hat {N}}}{dz^{2}}
+ \{ c - (a + b + 1)z \}\frac{d\Psi_{\hat {E} \hat {N}}}{dz}
- ab \Psi_{\hat {E} \hat {N}} = 0,
\end{equation}
where
\begin{eqnarray}
a & = & 1/2 + i \lambda + i (\hat {E} + \hat {N})/2, \nonumber \\
b & = & 1/2 - i \lambda + i (\hat {E} + \hat {N})/2, \nonumber \\
c & = & 1 + i \hat {E}.
\end{eqnarray}
The general solution is written in terms of Kummer's fundamental system of
solutions for the hypergeometric differential equation:
\begin{eqnarray}
 T_{\hat {E} \hat {N}}(z) & = &
c_{1} U_{\hat {E} \hat {N}} + c_{2} V_{\hat {E} \hat {N}}
\;\;\; \mbox{for $|z| < 1$}, \\
& = &
\bar {c}_{1} \bar {U}_{\hat {E} \hat {N}} + \bar {c}_{2} \bar {V}_{\hat {E}
\hat {N}}
\;\;\; \mbox{for $|z| > 1$},
\end{eqnarray}
where
\begin{eqnarray}
U_{\hat {E} \hat {N}} & = & z^{i \hat {E}/2} (1 - z)^{i \hat {N}/2} F(a, b, c;
z), \\
V_{\hat {E} \hat {N}} & = & z^{i \hat {E}/2} (1 - z)^{i \hat {N}/2}
			z^{1 - c} F(a - c + 1, b - c + 1, 2 - c; z), \\
\bar {U}_{\hat {E} \hat {N}} & = & z^{i \hat {E}/2} (1 - z)^{i \hat {N}/2}
			(-z)^{- b} F(b, b - c + 1, b -a +1; 1/z), \\
\bar {V}_{\hat {E} \hat {N}} & = & z^{i \hat {E}/2} (1 - z)^{i \hat {N}/2}
			(-z)^{- a} F(a, a - c + 1, a - b + 1; 1/z).
\end{eqnarray}
The modes $(U, V)$ are analogous to the Rindler modes. These modes obey
$V^{\lambda}_{\hat {E} \hat {N}} = (U^{-\lambda}_{\hat {E} -\hat {N}})^{*}$ and
$\bar {V}^{\lambda}_{\hat {E} \hat {N}} = (\bar {U}^{\lambda}_{-\hat {E} -\hat
{N}})^{*}$.  The relation between $ \bar {c}_{i}$ and $c_{i}$ are obtained
using the linear transformation properties of hypergeometric functions
\cite{abramowitz}:
\begin{eqnarray}
\bar {c}_{1} & = &
c_{1} \frac{\Gamma (c) \Gamma (a - b)}{\Gamma (a) \Gamma (c - b)}
+ c_{2} \frac{\Gamma (2 - c)\Gamma (a - b)}{\Gamma (a - c + 1)\Gamma (1 - b)},
	\\
\bar {c}_{2} & = &
c_{1} \frac{\Gamma (c) \Gamma (b - a)}{\Gamma (b) \Gamma (c - a)}
+ c_{2} \frac{\Gamma (2-c) \Gamma (b - a)}{\Gamma (b - c + 1)\Gamma (1 - a)}.
\end{eqnarray}
The constants $ c_{i} $ are determined by imposing the appropriate boundary
conditions below.

Near the horizon ($r \rightarrow r_{+} \; \mbox{or} \; z \rightarrow 0$), the
general solution approaches to
\begin{eqnarray}
T_{\hat {E}\hat {N}}(z \rightarrow 0) & \simeq &
c_{1} (-z)^{i \hat {E}/2} + c_{2} (-z)^{-i \hat {E}/2}
	\nonumber \\
& \simeq & c_{1} e^{i \hat {E} \ln \sqrt{r^{2}-r_{+}^{2}}}
	+ c_{2} e^{- i \hat {E} \ln \sqrt{r^{2}-r_{+}^{2}}}.
	\label{eq:zero}
\end{eqnarray}
Similarly, in the asymptotic region ($r \rightarrow \infty \; \mbox{or} \; z
\rightarrow -\infty$),
\begin{equation}
T_{\hat {E}\hat {N}}(z \rightarrow -\infty) \simeq
\bar {c}_{1} (-z)^{-1/2 + i \lambda} + \bar {c}_{2} (-z)^{-1/2 - i \lambda}.
	\label{eq:infty}
\end{equation}
Here, we omit various constants and phases which are irrelevant to later
discussion. Henceforth, we set $ E > 0 $ and $ \lambda > 0 $. Then, the first
and second terms in (\ref{eq:zero}) and (\ref{eq:infty}) represent outgoing and
ingoing modes respectively in the $s$-wave sector. Note that $ \lambda $, not $
E $, plays the role of the ``radial momentum" from the asymptotic behavior
(\ref{eq:infty}). This is because of the unusual asymptotic geometry; the
geometry is anti-de~Sitter rather than Minkowski.

First we investigate the tachyon scattering off the black hole. The appropriate
choice for the constants is $ c_{1} = 0 $. Asymptotically this solution has
both ingoing and outgoing modes, but only ingoing modes exist near the horizon:
\begin{equation}
T_{\hat {E} \hat {N}}(r \rightarrow r_{+})\simeq e^{-i\hat {E}\ln \sqrt
{r^{2}-r_{+}^{2}}},
\end{equation}
and
\begin{equation}
T_{\hat {E}\hat {N}}(r \rightarrow \infty) \simeq
\bar {c}_{1} e^{-(1-2i\lambda)\ln r} + \bar {c}_{2} e^{-(1+2i\lambda)\ln r}.
\end{equation}
Here, $c_{2}$ has been normalized to unity. The reflection coefficient is now
easily read as
\begin{eqnarray}
R & = &
\left| \frac{\Gamma (a-b) \Gamma (b-c+1) \Gamma (1-a)}{\Gamma (a-c+1) \Gamma
(1-b) \Gamma (b-a)} \right|^{2}	\nonumber \\
& = & \frac {\cosh \pi \left(\lambda - \frac{E + N}{2(r_{+} - r_{-})}\right)
\cosh \pi \left(\lambda - \frac{E - N}{2(r_{+} + r_{-})}\right)}
{\cosh \pi \left(\lambda + \frac{E + N}{2(r_{+} - r_{-})}\right)
\cosh \pi \left(\lambda + \frac{E - N}{2(r_{+} + r_{-})}\right)}.
	\label{eq:reflection}
\end{eqnarray}

In general, tachyon equations in rotating black hole backgrounds are more
complicated than the ones in non-rotating black hole backgrounds and may not be
easy to solve. The BTZ black hole case is special, because the transformation
(\ref{eq:transf}) relates the $J \neq 0$ metric to the $J = 0$
metric.\footnote{Of course, this does not mean that the rotating metric is the
same as the non-rotating metric because orbifoldings are different. For a
related issue, see ref.~\cite{DJtH}.}  This fact was crucial to solve the
tachyon equation (\ref{eq:nonzeroJ}). In fact, the eq.~(\ref{eq:zeroJ}) is
nothing but the tachyon equation for the static $ M=1 $ BTZ black hole. This
equation appeared in the study of string scattering in the static BTZ black
hole background \cite{ghoroku}.

This property, the $J \neq 0$ metric can be transformed to the $J = 0$ metric,
is special to the BTZ solution. Most of the other rotating black hole metrics
cannot be cast in the form of non-rotating black hole metrics. For example,
written in Boyer-Lindquist coordinates, the Kerr solution has the form
\begin{eqnarray}
ds^{2} & = &
	- \frac{ \Delta }{ \rho^2 } \{ dt - a \sin^2 \theta \, d\varphi \}^2
	+ \frac{ \sin^2 \theta }{ \rho^2 } \{ (r^2 + a^2) d\varphi - a \, dt \}^2
\nonumber \\
	&   & \mbox + \frac{ \rho^2 }{ \Delta } dr^2 + \rho^2 d\theta^2,
	\label{eq:kerr}
\end{eqnarray}
where
\begin{eqnarray}
\Delta & = & r^2 - 2Mr + a^2,
	\nonumber \\
\rho^2 & = & r^2 + a^2 \cos^2\theta,
\end{eqnarray}
and $a$ is the angular momentum per unit mass. This metric cannot be written in
the form of the Schwarzschild solution because the expressions inside curly
brackets in (\ref{eq:kerr}) are not integrable.

\section{Hawking Radiation}

A different choice for the constants $ c_{i} $ is usable to derive the Hawking
temperature.  Since the modes $(U, V)$ and $(\bar {U}, \bar {V})$ are related
by a Bogoliubov transformation, one has to simply determine the Bogoliubov
coefficients to get the Hawking temperature \cite{birrell}. By setting $\bar
{c}_{2} =1$ and
\begin{equation}
\bar {c}_{1} = c_{1}\frac{\Gamma (c) \Gamma  (a - b)}{\Gamma (a) \Gamma (c -
b)} + c_{2}\frac{\Gamma (2 - c) \Gamma  (a - b)}{\Gamma (a - c + 1) \Gamma (1 -
b)} = 0,
	\label{eq:bc}
\end{equation}
$c_{1}$ and $c_{2}$ become the Bogoliubov coefficients:
\begin{equation}
\bar {V}_{\hat {E} \hat {N}} =
c_{1} U_{\hat {E} \hat {N}} + c_{2} V_{\hat {E} \hat {N}}.
\end{equation}
Thus, the expectation value of the number operator $N_{E N}$ for $(U, V)$ mode
particles in the vacuum of the $(\bar {U}, \bar {V})$ mode $ \ket{\bar{0}} $ is
given by
\begin{equation}
\bra{\bar{0}} N_{E N} \ket{\bar{0}}
= \frac {|c_{1}|^2}{|c_{2}|^2 - |c_{1}|^2}.
\end{equation}
{}From (\ref{eq:bc}), we get
\begin{equation}
\bra{\bar{0}} N_{E N} \ket{\bar{0}}
= \frac {R}{1-R},
\end{equation}
where $R$ is the reflection coefficient in (\ref{eq:reflection}). In the limit
$ (E \pm N)/(r_{+} \pm r_{-}) \ll \lambda $, this expression reduces to
\begin{equation}
\bra{\bar{0}} N_{E N} \ket{\bar{0}} = \frac{1}{e^{(E - N \Omega)/T_{Hawk.}}-1},
	\label{eq:thermal}
\end{equation}
with
\begin{equation}
T_{Hawk.} = \frac{r_{+}^{2}-r_{-}^{2}}{2\pi r_{+}}, \;\;\;
\Omega = \frac {r_{+}}{r_{-}}.
\end{equation}
Since $ \Omega = J/2 r_{+}^2 $, $\Omega$ is the angular velocity of the
horizon. Eq.~(\ref{eq:thermal}) is the correct distribution function for the
rotating black hole with the Hawking temperature $T_{Hawk.}$ and the angular
velocity of the horizon $\Omega$. One can regard the limit $ (E \pm N)/(r_{+}
\pm r_{-}) \ll \lambda $ as the ``large radial momentum" limit because $
\lambda $ plays the role of the radial momentum as discussed in section~3.

The dependence $ E - N \Omega $ on (\ref{eq:thermal}) rather than $E$ is a sign
of rotating black holes. This represents the effect of the rotation and has
following consequences \cite{birrell}.

First, note that the effect of the rotation enters into the thermal spectrum in
the same way as a chemical potential. The angular momentum of the black hole
plays the role of a chemical potential. This factor (\ref{eq:thermal}) is
larger for positive $N$ than for negative $N$. Thus, it is favorable to emit
particles with angular momenta in the same direction as that of the black hole.
 As a result of the emission, the black hole rotation will be slowed down.

Also, (\ref{eq:thermal}) is negative when $ E < N \Omega $. This has an
interesting consequence. For simplicity, consider the limit $ T_{Hawk.}
\rightarrow 0 $. For static black holes where $ \Omega = 0 $, the Hawking
emission dies away in this limit; $ \bra{\bar{0}} N \ket{\bar{0}} \rightarrow
0$. However, for rotating black holes, $ \bra{\bar{0}} N \ket{\bar{0}} $ does
not die away when $ E < N \Omega $. In fact,  $ \bra{\bar{0}} N \ket{\bar{0}}
\rightarrow -1$. This negative flux means that the black hole induces
stimulated emission. This phenomenon is known as super-radiance.

\vspace{.1in}

\begin{center}
    {\Large {\bf Acknowledgements} }
\end{center}
\vspace{.1in}

We would like to thank K.~Ghoroku, A.~Ishibashi, and Y.~Satoh for useful
discussions. The work of M.~N. was supported in part by the Research Fellowship
of the Japan Society of the Promotion of Science for Young Scientists. The work
of N.~S. was supported in part by Grant-in-Aid for Scientific Research
No.~05640334 from the Ministry of Education, Science, and Culture.

\pagebreak

\end{document}